\documentclass[aps,pra,twocolumn,showpacs,floatfix]{revtex4}
\usepackage{epsfig}
\usepackage{graphicx}
\usepackage{dcolumn}
\usepackage{longtable}
\usepackage{amsthm,amsmath}

\begin{document}

\preprint{\today}

\title{Correlation trends in the ground state static electric dipole polarizabilities of closed-shell atoms and ions}

\author{Yashpal Singh\footnote{Email: yashpal@prl.res.in} and B. K. Sahoo \footnote{Email: bijaya@prl.res.in}}
\affiliation{Theoretical Physics Division, Physical Research Laboratory,
Navrangpura, Ahmedabad - 380009, India}

\author{B. P. Das}
\affiliation{Theoretical Physics and Astrophysics Group, Indian Institute of Astrophysics, Bangalore-560034, India}

\begin{abstract}
We employ the closed-shell perturbed relativistic coupled-cluster (RCC) theory developed by us earlier [Phys. Rev. A {\bf 77}, 062516 (2008)] to  
evaluate the ground state static electric dipole polarizabilities ($\alpha$s) of several atomic systems.
In this work, we have incorporated a class of higher order many-body effects in our
calculations that had not been taken into account in the above paper. We highlight their importance in improving the accuracies
of $\alpha$s. We also calculate the ground state $\alpha$s of the inert gas atoms and several
iso-electronic singly and doubly charged ions in order to make a comparative
studies of the trends of the correlation effects. Furthermore, we
have developed a method to construct intermediate diagrams that are required 
for the computations of the unperturbed singles and doubles coupled-cluster 
amplitudes. Our RCC results are compared with those of many-body perturbation theory
at different orders to demonstrate the importance of higher order correlation effects for the accurate 
determination of ($\alpha$s) of the systems that we have considered.
\end{abstract}

\pacs{31.15.ap, 31.15.bw, 31.15.ve, 31.15.xp}

\maketitle

\section{Introduction}
The subject of atom-light interaction has received considerable attention with the advent of 
sophisticated techniques to trap and cool atomic systems and measure their properties to very 
high precision \cite{foot}. An accurate knowledge of the electric dipole polarizabilities of the atomic
states are essential in these experiments as they are required in the studies of 
atomic interactions in optical lattices, atomic clocks, quantum information and many 
other important areas of atomic and molecular physics \cite{bonin,pethick,madej}. This range of applications
of electric dipole polarizabilities ($\alpha$) puts a premium on their accurate determination
in atomic systems. Precise measurements of $\alpha$ are challenging and involve
using a number of techniques like deflection of atomic beam by electric field \cite{hall}, E-H balance method
\cite{molof,miller,schwartz}, atom interferometry \cite{cronin,ekstrom}, cold atom 
velocity change \cite{amini} et cetera. In fact, the  ground state $\alpha$ of many atomic systems 
are not yet measured very precisely owing to difficulties in eliminating some of the larger
systematics. Therefore, accurate theoretical studies of electric dipole polarizabilities in 
atomic systems are of particular interest.
 
Dalgarno and his collaborators initiated work on the polarizabilities of many-electron atoms more than 
five decades ago \cite{dalgarno, dalgarno2}. Currently one of the most advanced
approaches to this property is  based on the linear response coupled cluster
theory \cite{monkhorst,daalgard,kundu,koch,kobayashi,datta,kowalski,hammond}. We had formulated a 
relativistic coupled-cluster (RCC) method to calculate polarizabilities in which the electric dipole operator was effectively treated 
as a perturbation \cite{bijaya1}. The first order perturbed RCC wave function in this case
was obtained by solving an inhomogeneous equation, thereby circumventing the sum-over-states 
approach \cite{bijaya1,bijaya2,bijaya3}. This method has been used to compute the polarizabilities of 
the ground states for a number of atomic systems \cite{sidhu-ne,sidhu-nobel,sidhu-alkali,sidhu-alkaline}.
In this work, we apply our method to some systems that we had studied earlier in addition to many
new candidates for acquiring insights into the behavior of electron correlation in the calculations
of the static electric dipole polarizabilities of these systems. We have included  higher order
non-linear RCC terms through special computational techniques, thereby improving our previous
results. We present our Dirac-Fock (DF), third order many-body
perturbation theory (MBPT(3)) and RCC results to show the quantitative changes in correlation effects in
the passage from finite order MBPT to RCC and we offer explanations for this behavior.

The rest of the paper is organized as follows: In the next section, we present briefly 
the theory of static electric dipole polarizability and the basic working equations for its
evaluation in the framework of the method that we have developed. Then, we describe the procedure
for obtaining the atomic wave function and the electric dipole polarizability using the RCC method.
This is followed by discussions of our results which are compared with the other calculations
and measurements. We use atomic units (au) in this paper. 

\section{Theory}
\subsection{Theory of static dipole polarizability}
The second order change in the energy of an atomic state $|\Psi_n^{(0)} \rangle$ when placed in an
external weak electric field {\bf E$={\cal E} \hat r$} is given by
\begin{equation}
 \delta E = -\frac{1}{2}\alpha {\cal E}^2
\label{eq1}
\end{equation}
where $\alpha$ is known as the static electric dipole polarizability of the state which can be
mathematically expressed using the second order perturbation theory as
\begin{equation}
 \alpha=-\frac{2}{\langle \Psi_n^{(0)}| \Psi_n^{(0)} \rangle } \sum_{I}\frac{|\langle \Psi_n^{(0)}|D|\Psi_I^{(0)} \rangle|^2}{E_n^{(0)}-E_I^{(0)}},
\label{eq2}
\end{equation}
where the summation over $I$ represents the inclusion of all possible intermediate 
states $|\Psi_I^{(0)} \rangle$ and $E^{(0)}$s are the energies of the respective
states denoted by the index in the subscripts.
It is possible to determine $\alpha$ by calculating the second order derivative of the energy shift given by Eq. (\ref{eq1}) with
respect to the electric field strength $\cal{E}$. A straightforward approach 
would be to sum over the intermediate states explicitly in Eq. (\ref{eq2}). However, this is not very practical from a numerical point of view.
In the method that we used earlier, polarizabilities can be determined without summing over intermediate states.
In that approach, we express Eq. (\ref{eq2}) as
\begin{equation}
 \alpha=- 2 \frac{\langle \Psi_n^{(0)}|D|\Psi_n^{(1)} \rangle}{ \langle \Psi_n^{(0)}| \Psi_n^{(0)} \rangle },
\label{eq3}
\end{equation}
where $|\Psi_n^{(1)} \rangle$ is a modified wave function similar to the first order
perturbed wave function of $|\Psi_n^{(0)} \rangle$ which is given by
\begin{eqnarray}
|\Psi_n^{(1)} \rangle &=& \sum_{I}|\Psi_I^{(0)} \rangle \frac{|\langle \Psi_I^{(0)}|D|\Psi_n^{(0)} \rangle}{E_n^{(0)}-E_I^{(0)}}. \ \ \ \
\label{eq4}
\end{eqnarray}
It is clear that $|\Psi_n^{(1)} \rangle$ is obtained by perturbing 
$|\Psi_n^{(0)} \rangle$ by the vector operator $D$. Therefore, $|\Psi_n^{(1)} \rangle$ can be obtained by
solving the inhomogeneous equation
\begin{eqnarray}
(H-E_n) |\Psi_n^{(1)} \rangle &=& -D|\Psi_n^{(0)} \rangle.
\label{eq5}
\end{eqnarray}
This procedure is similar in spirit to the approach given in \cite{dalgarno} in the sense that our first order perturbed wave function
is obtained from an inhomogeneous equation rather than by summing over intermediate states. To be specific, in the present
work $|\Psi_n^{(1)} \rangle$ is the first order correction to the atomic wave function with the electric dipole operator
as the perturbation. Now expressing the total wave function as $|\Psi_n \rangle = |\Psi_n^{(0)} \rangle 
+ \lambda |\Psi_n^{(1)} \rangle$ and owing to the fact that $D$ is an odd hermitian operator, we have
\begin{equation}
 \alpha= \frac{\langle \Psi_n|D|\Psi_n \rangle}{\langle \Psi_n|\Psi_n \rangle},
\label{eq6}
\end{equation}
from keeping terms up to linear in the arbitrary parameter $\lambda$.

\subsection{Generalized Bloch equation for two external perturbations}
The generalized Bloch equation for the model Hamiltonian $H_0$ and the interaction term $V$
is given by \cite{lindgren}
\begin{eqnarray}
[\Omega,H_0]P &=& QV \Omega P - \chi P V \Omega P,
\label{eq7}
\end{eqnarray}
where $P$ and $Q$ are the projection operators corresponding to the model and orthogonal spaces respectively and $\Omega = 1 + \chi$ 
is the wave operator that generates the atomic state function from the reference state $|\Phi_n \rangle$
of a Hamiltonian say $H_0$; i.e. $|\Psi_n \rangle = \Omega |\Phi_n \rangle$. By expanding the wave operator order by order,
we obtain the following working equation
\begin{eqnarray}
 [\Omega^{(k)},H_0 ]P=QV\Omega^{(k-1)}P - \sum_{m=1 }^{k-1}
\nonumber \Omega^{(k-m)}PV\Omega^{(m-1)}P , \\
\label{eq8}
\end{eqnarray}
where $\Omega^{(0)}=1$ and the superscript $k$ represents orders of $V$ present in the evaluation of $\Omega^{(k)}$. When there are
two sources of perturbation, we can still express $|\Psi_n \rangle = \Omega |\Phi_n \rangle$ with the new perturbation
potential $V=V_1 + V_2$. In this case, the $k^{th}$ order $\Omega^{(k)}$ is redefined by $\Omega^{(\beta,\delta)}$
such that $k=\beta + \delta$ for the $\beta$ orders of $V_1$ and $\delta$ orders of $V_2$. In this case, the corresponding 
Bloch equation is expressed by \cite{bijaya}
\begin{eqnarray}
 [\Omega^{(\beta,\delta)},H_0 ]P &=& QV_1 \Omega^{(\beta-1,\delta)}P + Q V_2 \Omega^{(\beta,\delta-1)}P -
\nonumber \\ && \sum_{m=1 }^{\beta-1}
 \sum_{l=1}^{\delta-1} \big ( \Omega^{(\beta-m,\delta-l)}
 P V_1 \Omega^{(m-1,l)}P - \nonumber \\ && \Omega^{(\beta-m,\delta-l)}PV_2 \Omega^{(m,l-1)}P \big ),
\label{eq9}
\end{eqnarray}
with $\Omega^{(0,0)}=1$, $\Omega^{(1,0)}=V_1$ and $\Omega^{(0,1)}=V_2$. In this procedure, the
atomic state function up to $k^{th}$ order is given by
\begin{eqnarray}
|\Psi_n^{(k)} \rangle = [ \Omega^{(k,0)} + \sum_{\delta=1}^{k-1} \lambda^{\delta} \Omega^{(k-\delta, \delta)} ] |\Phi_n \rangle,
\label{eq10}
\end{eqnarray}
where we have introduced a parameter $\lambda$ without any loss of generality to keep track of the order of $V_2$
and it can be later set to one in the final consideration.

Using the above prescription in Eqs. (\ref{eq5}) and (\ref{eq3}), it yields
\begin{equation}
|\Psi_n^{(k,0)} \rangle = \Omega^{(k,0)} |\Phi_n \rangle
\label{eq11}
\end{equation}
and its first order correction due to $V_2$ is given by
\begin{equation}
|\Psi_n^{(k,1)} \rangle = \Omega^{(k,1)} |\Phi_n \rangle .
\label{eq12}
\end{equation}

In our present work, the residual interaction $V_{es}$ is treated as the first
perturbation ($V_1=V_{es}$) to include the electron correlation effects in a many-body 
perturbation treatment and we set $V_2=D$ and $\delta=1$ for calculating $\alpha$. 
In this approach, the lowest and $k^{th}$ order results for $\alpha$ is given by
\begin{equation}
 \alpha=2 \langle \Psi_n|D \Omega^{(0,1)} |\Psi_n \rangle.
\label{eq13}
\end{equation}
and
\begin{equation}
\alpha=2 \frac{\sum_{\beta=0}^{k-1} \langle \Phi_n| {\Omega^{(k-\beta-1,0)}}^{\dagger} D \Omega^{(\beta,1)} |\Phi_n \rangle}
{ \sum_{\beta=0}^{k-1} \langle \Phi_n| {\Omega^{(k-\beta-1,0)}}^{\dagger} \Omega^{(\beta,0)} |\Phi_n \rangle},
\label{eq14}
\end{equation}
respectively.

\begin{figure}[t]
\includegraphics[width=9.0cm]{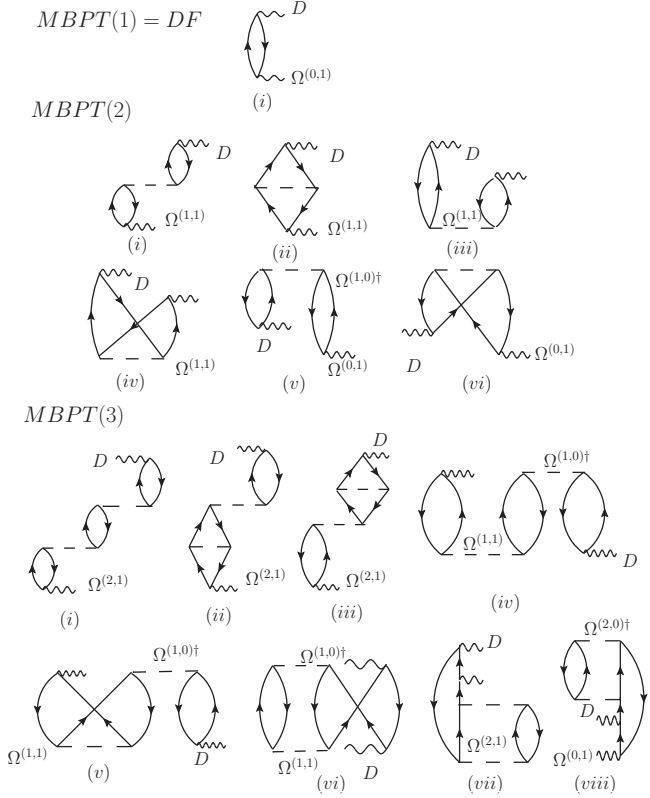}
\caption{Few important contributing diagrams of the MBPT(3) method. The lowest order contribution 
is given as the DF result.}
\label{fig1}
\end{figure}

\section{Method of calculations}
The Dirac-Coulomb (DC) atomic Hamiltonian which is used in our calculation is given by
\begin{eqnarray}
H^{DC} &=&\sum_i \left [ c\mbox{\boldmath$\alpha$}_i\cdot \textbf{p}_i+(\beta_i -1)c^2+
V_{nuc}(r_i) + \sum_{j>i} \frac{1}{r_{ij}} \right ] \nonumber \\
\label{eq15}
\end{eqnarray}
The single particle energies are scaled with respect to the rest mass energy of the electron, the
nuclear potential is evaluated considering a Fermi nuclear charge distribution and the
electron-electron interaction due to one photon exchange is restricted to Coulomb
interactions only.

The DF approximation ($H_0=H_{DF}$) yields
the mean-field wave function $|\Phi_0 \rangle$ for the ground state which we consider as the 
reference state $|\Phi_n \rangle$ for the construction of exact ground state  
wave function $|\Psi_n \rangle=|\Psi_0 \rangle$.

\begin{figure}[t]
\label{effsingles}
\label{one-body}
\includegraphics[width=8cm]{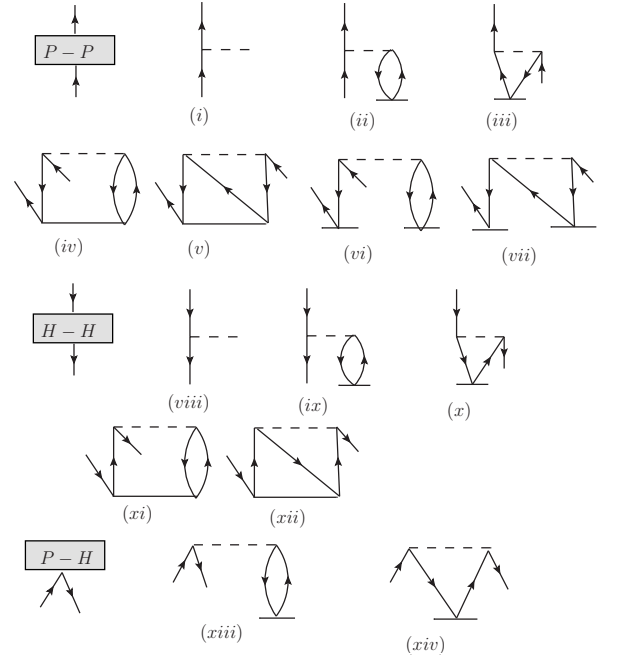}
\caption{Effective one-body intermediate diagrams for the evaluation of CCSD amplitudes. }
\label{efsng}
\end{figure}

\begin{figure}[t]
\label{effdoubles}
\centering
\includegraphics[width=8cm]{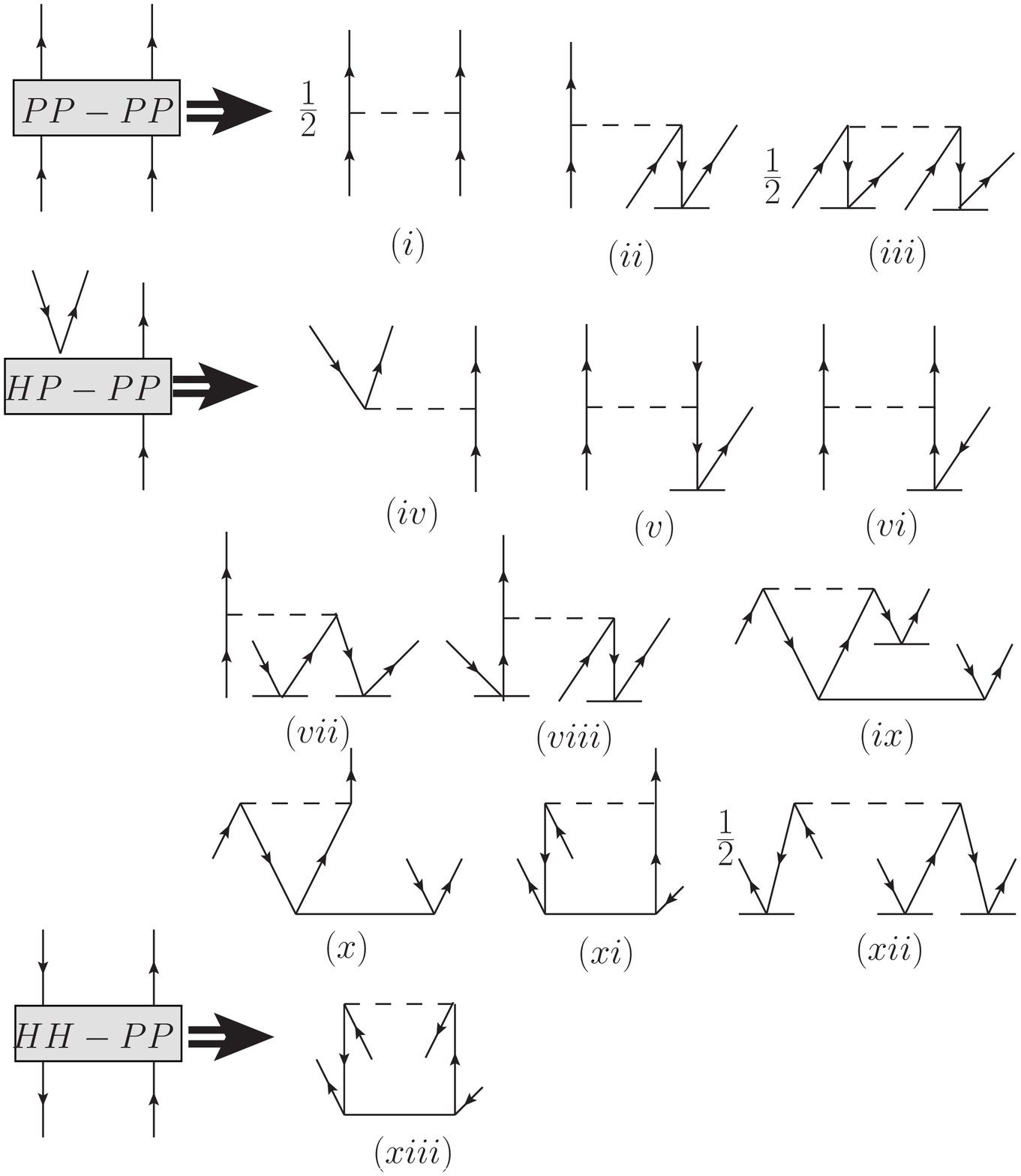}
\includegraphics[width=9cm]{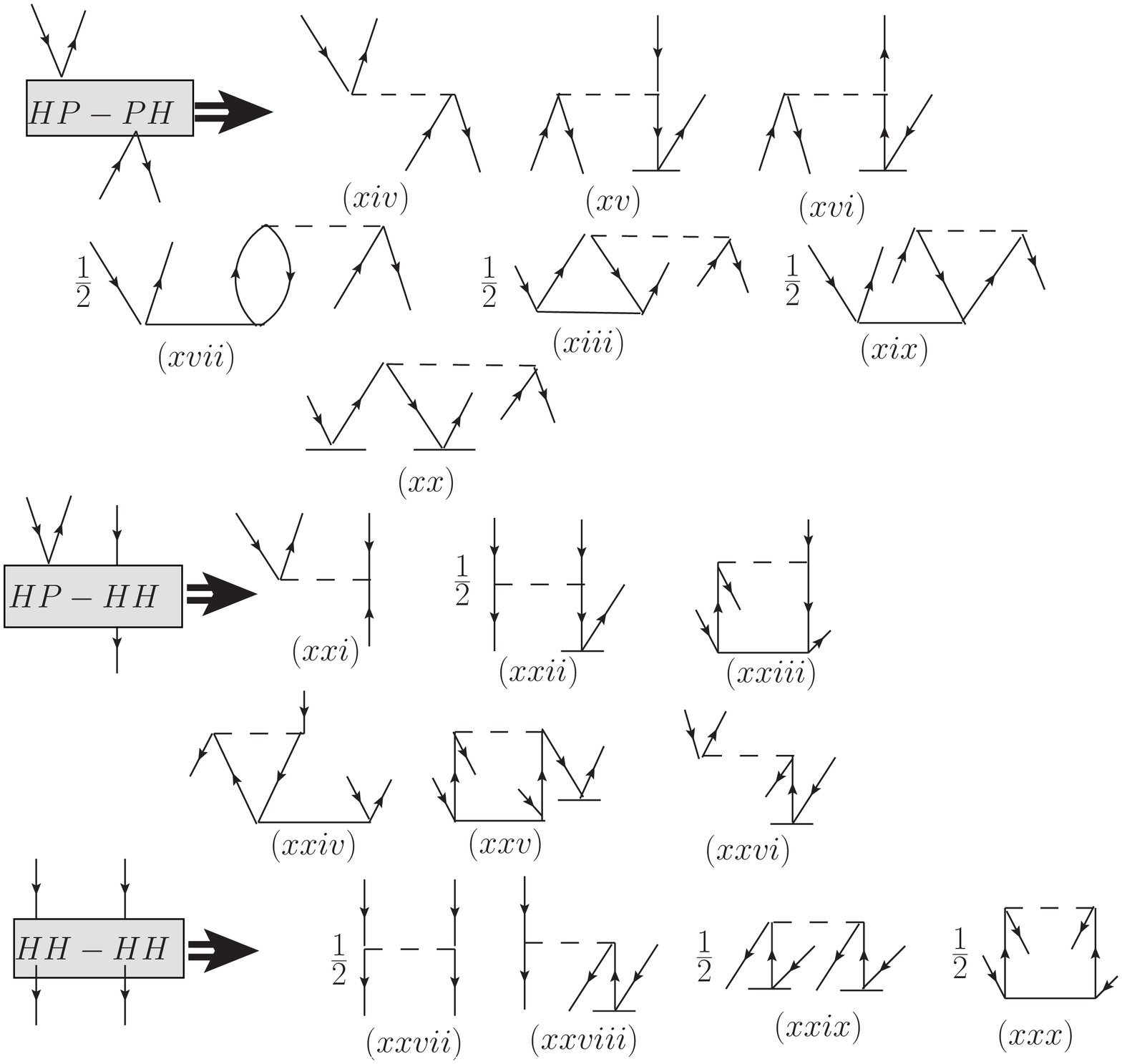}
\caption{Effective two-body intermediate diagrams for the evaluation of CCSD amplitudes. }
\label{efdb}
\end{figure}

\subsection{MBPT(3) method}
The importance of various correlation terms in the determination of $\alpha$ can be better understood
from an explicit analysis of lower order perturbation calculations where the contributions from the individual
term can be found explicitly. For this purpose, we have considered up to $\beta=2$ and $\delta=1$ 
(MBPT(3) method) to calculate $\alpha$.

Therefore, the expression for $\alpha$ of the ground state in the MBPT(3) method is given by
\begin{eqnarray}
\alpha &=& \frac{2}{\cal N} \langle \Phi_0 | [\Omega^{(0,0)}+\Omega^{(1,0)}+\Omega^{(2,0)}]^{\dagger} D \nonumber \\ && \times[\Omega^{(0,1)}+\Omega^{(1,1)}+\Omega^{(2,1)}]|\Phi_0 \rangle \nonumber \\
&=& \frac{2}{\cal N} \langle \Phi_0| D\Omega^{(0,1)} + D\Omega^{(1,1)}+D\Omega^{(2,1)} + {\Omega^{(1,0)}}^{\dagger} D\Omega^{(0,1)}  \nonumber \\ && +
{\Omega^{(1,0)}}^{\dagger} D\Omega^{(1,1)} +{\Omega^{(2,0)}}^{\dagger} D\Omega^{(0,1)}|\Phi_0 \rangle  ,
\label{eq16}
\end{eqnarray} 
with ${\cal N}=\langle \Phi_0| 1 + \Omega^{(1,0)}+\Omega^{(2,0)} + {\Omega^{(1,0)}}^{\dagger} + {\Omega^{(2,0)}}^{\dagger} +{\Omega^{(1,0)}}^{\dagger} \Omega^{(0,1)} |\Phi_0 \rangle$.

The above wave operators are obtained using the following Bloch equations 
\begin{eqnarray} 
\nonumber [\Omega^{(1,0)},H_0]P&=& Q V_{es}P \\
\nonumber [\Omega^{(2,0)},H_0]P&=& Q V_{es} \Omega^{(1,0)}P - \Omega^{(1,0)}P V_{es} P \\
\nonumber \text{and} && \\
\nonumber[\Omega^{(0,1)},H_0]P&=&Q D P \\
\nonumber[\Omega^{(1,1)},H_0]P&=& Q V_{es} \Omega^{(0,1)}P + Q D \Omega^{(1,0)}P \\
\nonumber[\Omega^{(2,1)},H_0]P&=&Q V_{es}\Omega^{(1,1)}P + Q D \Omega^{(2,0)}P - \nonumber \\ && \Omega^{(1,0)}P V_{es} \Omega^{(0,1)} P 
 - \Omega^{(1,0)}P D \Omega^{(1,0)}P . \nonumber
\label{eq17}
\end{eqnarray}

Important lower order diagrams that contribute at the MBPT(3) level are shown in Fig. \ref{fig1}.

\begin{figure}[h]
\label{singles-con}
\includegraphics[width=6cm]{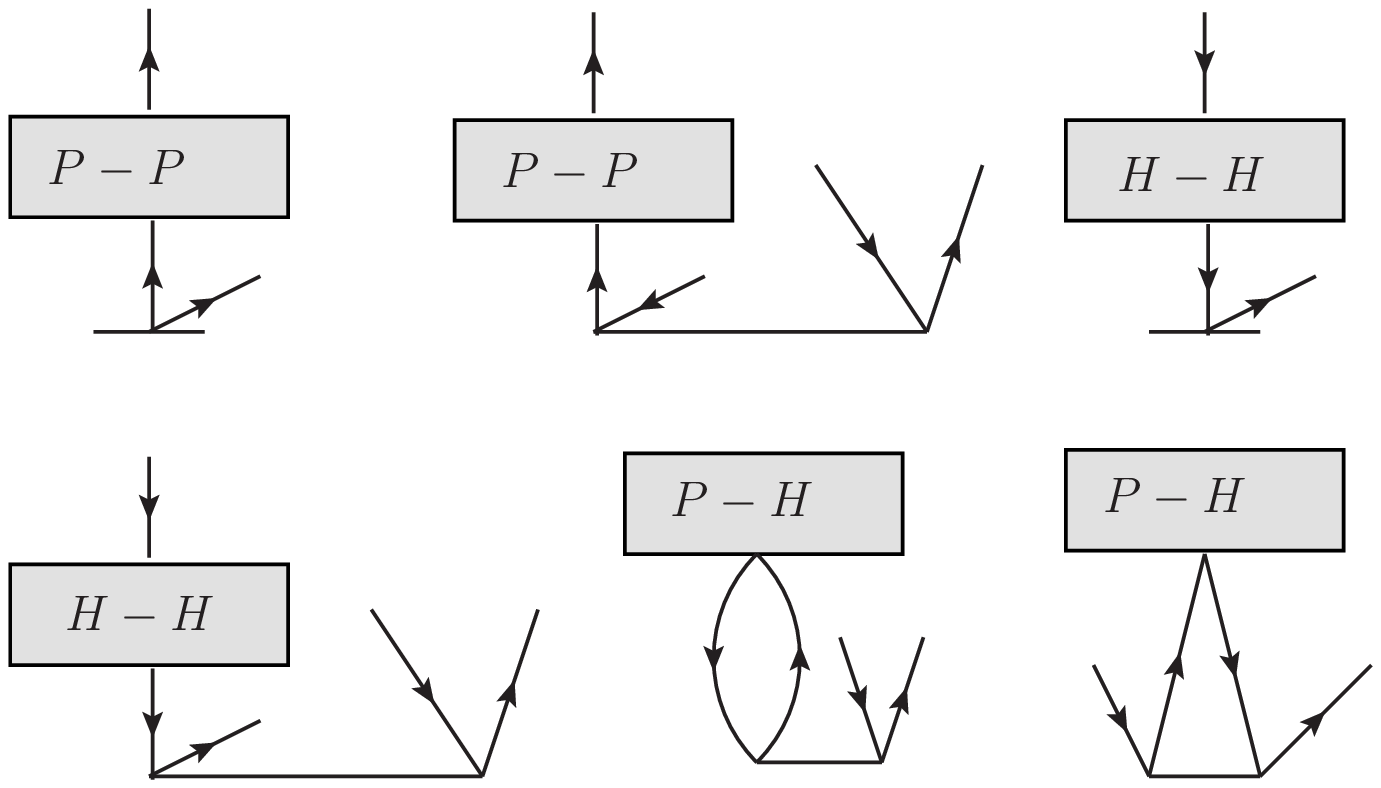}
\caption{Final CCSD amplitudes determining diagrams after contracting effective one-body intermediate diagrams with the $T^{(0)}$ operators.}
\label{ccd1}
\end{figure}
  
\begin{figure}[h]
\label{double-con}
\includegraphics[width=8cm]{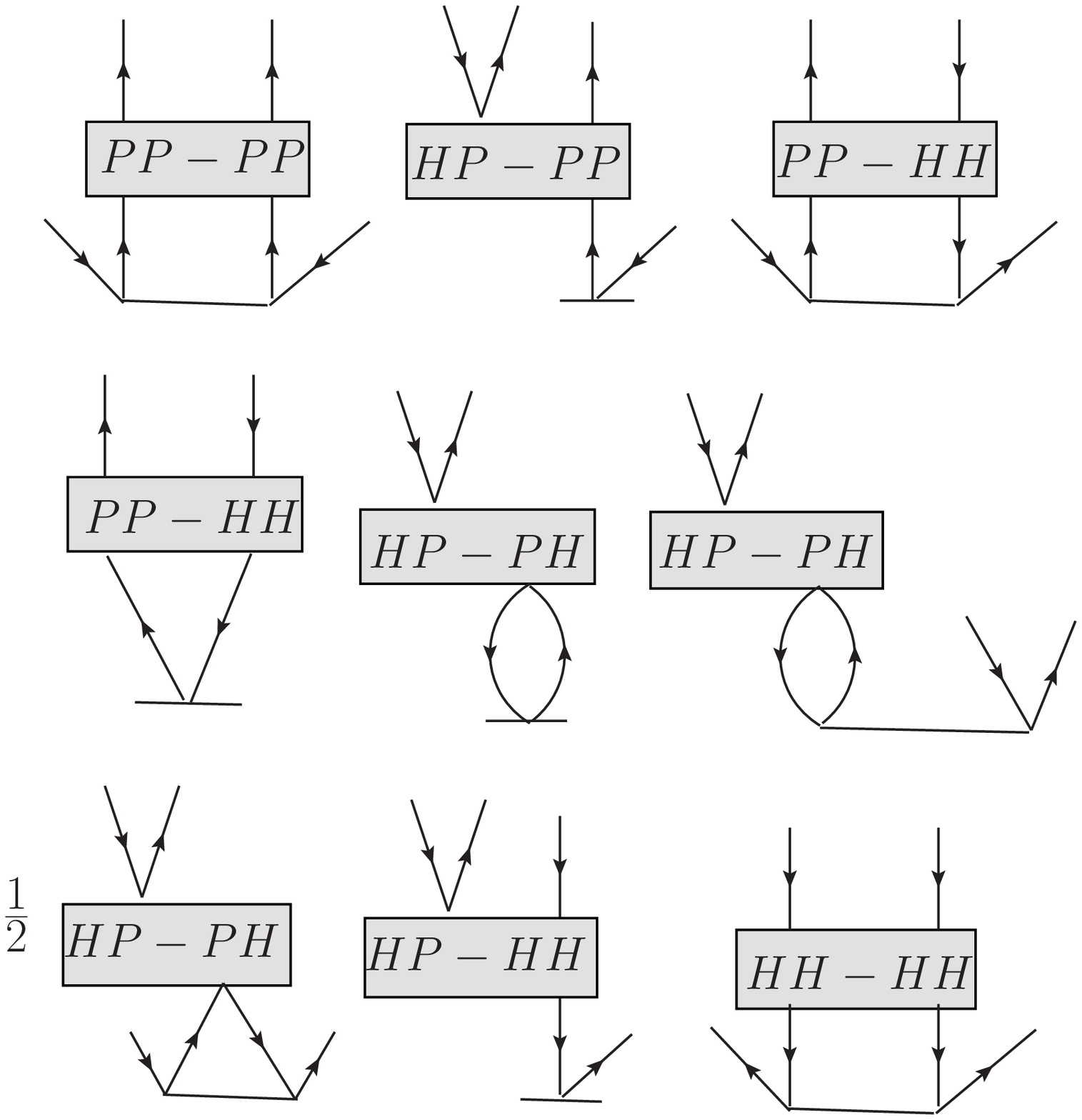}
\caption{Final CCSD amplitudes determining diagrams after contracting effective two-body intermediate diagrams with the $T^{(0)}$ operators.}
\label{ccd2}
\end{figure}
 
\subsection{RCC theory: CCSD method}

In the RCC theory, the ground state of a closed-shell atom is expressed as
\begin{eqnarray}
 | \Psi_0 \rangle &=& \Omega |\Phi_0 \rangle \nonumber \\
                  &=& e^T |\Phi_0 \rangle,
\label{eq18}
\label{eqcc}
\end{eqnarray}
where the operator $T$ corresponds to the excitations from the reference state $|\Phi_0 \rangle$. 

Following Eq. (\ref{eq6}), we have
\begin{eqnarray}
 \alpha &=& \frac{\langle \Phi_0| \Omega^{\dagger} D \Omega |\Phi_0 \rangle}{\langle \Phi_0| \Omega^{\dagger} \Omega | \Phi_0 \rangle} \nonumber \\
  &=& \frac{\langle \Phi_0| {e^T}^{\dagger} D e^T |\Phi_0 \rangle}{\langle \Phi_0| {e^T}^{\dagger} e^T | \Phi_0 \rangle }.
\label{eq19}
\end{eqnarray}

By definition, the $T$ operators are in normal order form with respect to the reference state $|\Phi_0 \rangle$. Therefore,
the above expression yields \cite{cizek}
\begin{eqnarray}
\alpha &=& \frac{\langle\Phi_0 |e^{T^\dagger}D_Ne^{T}|\Phi_0 \rangle_{con} \langle\Phi_0|e^{T^\dagger}e^{T}|\Phi_0\rangle}
{\langle\Phi_0|e^{T^\dagger}e^{T}|\Phi_0\rangle} \nonumber \\
&=&\langle\Phi_0 |e^{T^\dagger}D_Ne^{T}|\Phi_0 \rangle_{con} ,
\label{eq20}
\end{eqnarray}
this is a favorable denouement for the calculation of properties of the ground state of closed-shell atomic
systems using the RCC theory. The subscript $N$ represents the normal order form of $D$ and $con$ refers
to survival of only the connected diagrams.

\begin{figure}[t]
\label{sing-dc}
\includegraphics[width=8cm]{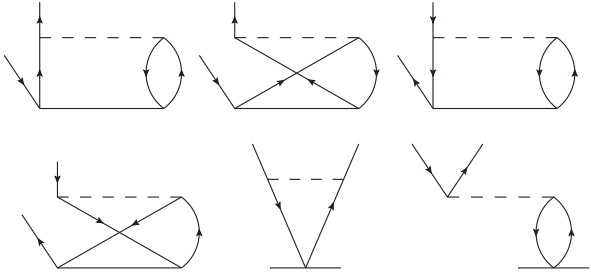}
\caption{Direct contributing diagrams to the singles of the CCSD method.}
\label{ccd3}
\end{figure}

\begin{figure}[t]
\label{doubles-dc}
\includegraphics[width=8cm]{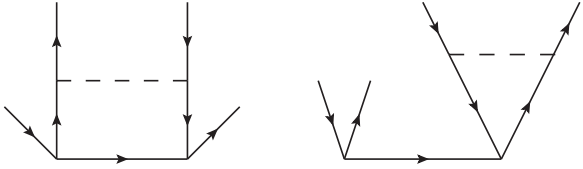}
\caption{Direct contributing diagrams to the doubles of the CCSD method.}
\label{ccd4}
\end{figure}
 
In the above RCC expression, the operator $T$ includes contributions from both $V_{es}$ and $D$. Taking this into consideration, we split $T$ into
\begin{equation}
 T=T^{(0)}+\lambda T^{(1)}
\label{eq21}
\end{equation}
where $T^{(0)}$ corresponds to correlation effects due to $V_{es}$ and $T^{(1)}$ takes into account the opposite parity excitations in the wave
function due to $D$. Substituting Eq. (\ref{eq21}) in Eq. (\ref{eq18}), we get
\begin{eqnarray}
 | \Psi_0^{(0)} \rangle &=& e^{T^{(0)}} |\Phi_0 \rangle 
\label{eq22}
\end{eqnarray}
and
\begin{eqnarray}
 | \Psi_0^{(1)} \rangle &=& e^{T^{(0)}} T^{(1)} |\Phi_0 \rangle. 
\label{eq23}
\end{eqnarray} 
In the present work, we have considered all possible singly and doubly excited configurations
(known as the CCSD method) by defining
\begin{eqnarray}
T^{(0)} &=& T_1^{(0)} + T_2^{(0)} 
\label{eq24}
\end{eqnarray}
and
\begin{eqnarray}
T^{(1)} &=& T_1^{(1)} + T_2^{(1)} .
\label{eq25}
\end{eqnarray}

\begin{figure}[t]
\includegraphics[width=8.5cm, height=8.5cm]{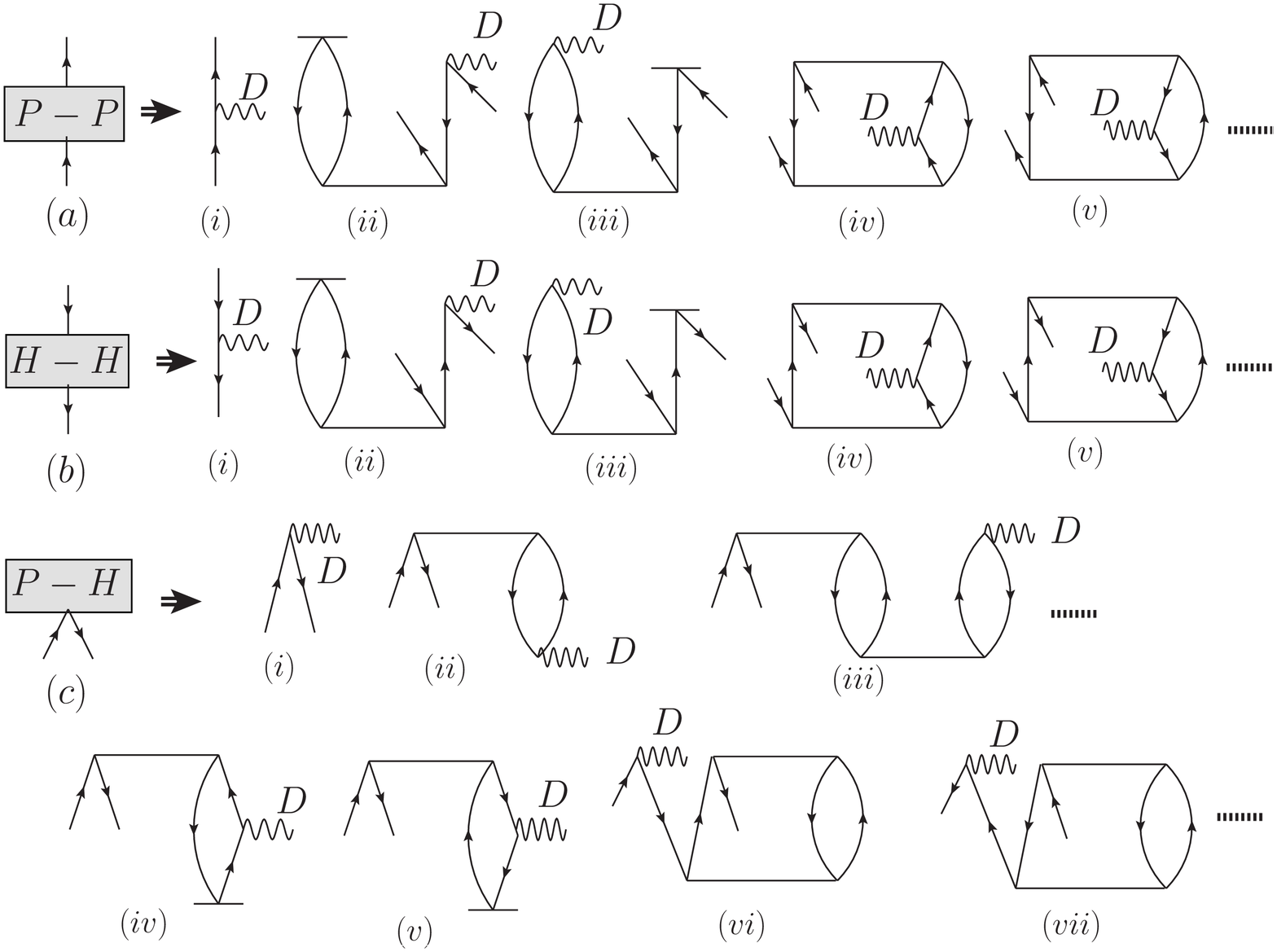}
\caption{Few effective one-body diagrams from $e^{T^{\dagger{(0)}}}D_Ne^{T^{(0)}}$ that are connected further with $T^{(1)}$ operator in the final evaluation of the polarizabilities.}
\label{prpf0}
 \end{figure}
\begin{figure}[t]
\includegraphics[width=8cm]{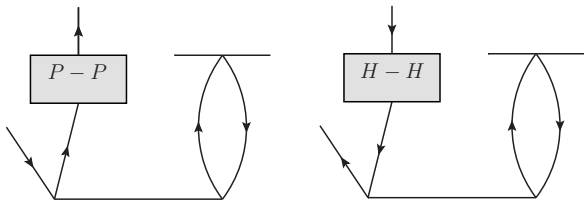}
\caption{Examples of further dressed-up effective H-P type diagrams from the effective P-P and H-H type one-body diagrams.} 
\label{prpf1}
 \end{figure}
The $T^{(0)}$ amplitudes are obtained by solving the following equation
\begin{equation}
 \langle \Phi_0^{\tau}|\overline{H_N^{DC}}|\Phi_0\rangle=0 ,
\end{equation}
where $\tau=1,2$ stands for either singly or doubly excited configurations from $|\Phi_0 \rangle$, subscript 
$N$ represents the normal ordered form of the DC Hamiltonian and the dressed Hamiltonian 
$\overline{H_N^{DC}}=e^{-T^{(0)}}H_N^{(DC)}e^{T^{(0)}}$ which is equal to $(H_N^{(DC)}e^{T^{(0)}})_{con}$ \cite{kavasnika}. 
To solve the above equation, we adopt the Jacobi iterative procedure and the non-linear terms from $(H_N^{(DC)}e^{T^{(0)}})_{con}$ 
are accounted through the intermediate diagrams. A standard procedure for defining intermediate diagrams are 
described in \cite{bartllet, harris}. However, we follow a different strategy here. The idea behind this is to avoid the  
repetition of defining different variables to store the intermediate diagrams. In our approach, we define some
distinct types of intermediate diagrams by classifying them into effective one-body and two-body diagrams 
as given in Figs.  \ref{efsng} and \ref{efdb}, respectively. They are further combined with 
suitable $T^{(0)}$ operators for constructing the final diagrams that contribute to the computations
of the $T_1^{(0)}$ and $T_2^{(0)}$ amplitudes as shown in Figs. \ref{ccd1} and \ref{ccd2}. The diagrams which 
involve fewer internal lines but are problematic for generating repetitive diagrams when included in the 
above effective one-body and two-body intermediate diagrams are computed directly. Examples of such 
diagrams for the evaluation of the $T_1^{(0)}$ and $T_2^{(0)}$ amplitudes are shown in Figs. \ref{ccd3} and 
\ref{ccd4}, respectively. To avoid the double counting of topologically equivalent diagrams arising from the
above effective intermediate diagrams and from symmetry considerations, we
multiply by a factor of $1/2$ wherever necessary as mentioned in Figs \ref{efdb} and \ref{ccd2}.

After obtaining $T^{(0)}$ amplitudes, we solve the following equation to determine the $T^{(1)}$ amplitudes 
\begin{eqnarray}
 \langle \Phi_0^{\tau}|\overline{H_N^{DC}}T^{(1)}|\Phi_0\rangle=-\langle \Phi_0^{\tau}|\overline{H_{int}}|\Phi_0\rangle .
\end{eqnarray}
We now make use of the locations of the previously defined intermediate variables to store all possible effective
one-body and two-body diagrams of $\overline{H_N^{DC}}$. In this case, no additional multiplicative factors are needed
owing to the fact that finally these diagrams are contracted with $T^{(1)}$ amplitudes. Also, all the
contributing terms from $T^{(0)}$ are included at this stage. Contributions from 
$\overline{H_{int}}=(H_{int}e^{T^{(0)}})_{con}$ are computed directly. 
\begin{table*}[t]
\caption{\label{tab1}
Ground state static dipole polarizability $\alpha$ of various closed shell atoms and ions. In the parentheses of our results, 
we have given the estimated uncertainties from the calculations. The square brackets refers to the references of other works.}
\begin{ruledtabular}
\begin{tabular}{lccc}
 Systems       &   This Work             &  Others          & Experiments\\
\hline                                                                                      \\
He    &1.360(20) & 1.322 \cite{johnson1}, 1.383763 \cite{soldan}     & 1.383223(67) \cite{gugan,gugan1}, 1.3838 \cite{langhoff}  \\
      &           &1.38376079(23) \cite{lach}, 1.382(1) \cite{bijaya1}    & 1.384 \cite{dalgarno1}, 1.383759(13) \cite{schmidt}              \\                              
Ne    & 2.652(15) &2.38 \cite{johnson1}, 2.6648 \cite{soldan}, 2.697 \cite{nakajima} & 2.670(3) \cite{orcutt}, 2.66110(3) \cite{gaiser}  \\
      &           &2.665 \cite{hald}, 2.668(6) \cite{franke} 2.6695 \cite{sidhu-ne}  & 2.6680 \cite{langhoff}, 2.663 \cite{dalgarno1}  \\
               
Ar     & 11.089(4)&10.77 \cite{johnson1}, 11.084 \cite{soldan}, 11.22 \cite{nakajima} & 11.081(5) \cite{orcutt}, 11.091 \cite{langhoff}   \\
       &          &11.085(6) \cite{lupinetti} 11.213 \cite{sidhu-nobel}& 11.080 \cite{dalgarno1}, 11.083(2) \cite{newell}  \\             
Kr       &  16.93(5)&  16.47 \cite{johnson1}, 16.80 \cite{thakkar},  16.736 \cite{sidhu-nobel}  & 16.766(8) \cite{orcutt}, 16.740  \cite{langhoff}, 16.740 \cite{dalgarno1} \\              
               
Be    & 37.86(17) & 37.755 \cite{komasa}, 37.73(5) \cite{tunega}, 37.807 \cite{bendazzoli}, 37.29 \cite{muller},             &             \\
      &      & 37.69 \cite{mitroy2}, 37.76 \cite{porsev1}, 37.80(47) \cite{bijaya1}       &               \\              
Mg      &  72.54(50) &71.7 \cite{archibong}, 70.90 \cite{hamonou}, 70.74 \cite{muller}, 71.35 \cite{mitroy2},                 &             \\
            &           & 71.33 \cite{porsev1}, 74.9(2.7) \cite{reshetnikov}, 73.41(2.32) \cite{bijaya1}                  &               \\                                                           
Ca     & 157.03(80)& 157 \cite{archibong}, 171.7 \cite{glass}, 156.0 \cite{muller}              & 169(17) \cite{miller}            \\
         &           & 159.4 \cite{mitroy2}, 158.00 \cite{lim}, 152 \cite{sadlej}               &               \\     
 &        &  159.0 \cite{porsev1}, 157.1(1.3) \cite{porsev1},  154.58(5.42) \cite{bijaya1}      &               \\ 
Sr    & 186.98(85) &201.2 \cite{mitroy2}, 198.85 \cite{lim}, 190 \cite{sadlej},   & 186(15) \cite{schwartz}            \\
     &           & 202.0  \cite{porsev1}, 197.2(2) \cite{porsev1}, 199.71(7.28) \cite{bijaya1} &               \\              
Li$^+$   & 0.1913(5) &0.192486 \cite{bhatia,johnson2}, 0.1894 \cite{johnson1}  &  0.1883(20) \cite{cooke}            \\                                                                               

Na$^+$    & 0.9984(7) & 0.9947 \cite{muller}, 0.9457 \cite{johnson1}& 0.978(10) \cite{opik}, 1.0015(15 \cite{freeman} \\
       &           & 1.00(4) \cite{lim2}, 1.025 \cite{sidhu-alkali}       & 0.9980(33 \cite{gray}               \\ 
K$^+$       & 5.522(7) &5.354 \cite{muller}, 5.457 \cite{johnson1}            & 5.47(5) \cite{opik}            \\
            &           &5.52(4) \cite{lim2}, 5.735 \cite{sidhu-alkali}          &               \\             
Rb$^+$     & 9.213(15) &9.076 \cite{johnson1}, 9.11(4) \cite{lim2}, 9.305\cite{sidhu-alkali}& 9.0 \cite{johansson}            \\               
Sc$^{+}$    & 53.24(20)          &          &             \\
       
Y$^{+}$   & 72.26(50)            &          &             \\
Be$^{2+}$   & 0.0521(2) & 0.05182 \cite{johnson1}, 0.052264 \cite{bhatia,johnson2} &             \\
Mg$^{2+}$    & 0.4852(5)            &0.4698 \cite{johansson}, 0.495 \cite{sidhu-alkaline}          & 0.489(5) \cite{opik}                 \\
            &           &0.4814 \cite{muller}             & 0.486(7) \cite{bockasten}              \\
             
Ca$^{2+}$    &   3.295(6) &  3.254 \cite{johnson1}, 3.161 \cite{muller}           & 3.26(3) \cite{opik}            \\
                 &           &  3.262 \cite{lim}, 3.387 \cite{sidhu-alkaline}           &               \\               
 
Sr$^{2+}$    &  5.877(8) &5.813 \cite{johnson1}, 5.792 \cite{lim}, 5.913 \cite{sidhu-alkaline}          &             \\
\end{tabular}
\end{ruledtabular} 
\end{table*}                                                                              
\begin{figure}[t]
\includegraphics[width=8cm]{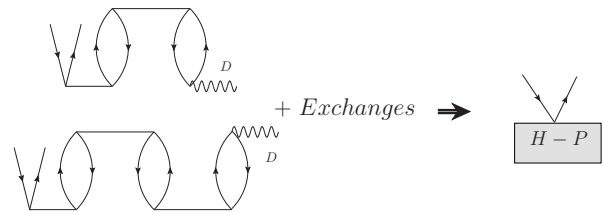}
\caption{Additional effective diagrams which are stored separately and contracted finally with 
the $T^{(0)} T^{(1)}$ for accounting higher order core-polarization correlation effects.}
\label{prpf2}
\end{figure}

\begin{table}[t]
\caption{\label{tabx}
Contributions to the estimated uncertainties from the Breit interaction ($\delta_B$), QED 
corrections ($\delta_Q$) and finite basis size ($\delta_F$) in the considered atomic systems.}
\begin{ruledtabular}
\begin{tabular}{lccc}
\textrm{System}& $\delta_B$ & $\delta_Q$ & $\delta_F$ \\
\hline                                                                                      \\
He        & $\sim 0.0$ & $\sim 0.0$ & 0.0200 \\
Ne        & 0.0006 & $\sim 0.0$ & 0.1499 \\
Ar        & 0.0023 & 0.0002 & 0.0032 \\
Kr        & 0.0148 & 0.0008 & 0.0478 \\
Be        & 0.0027 & $-0.0001$ & 0.1699 \\
Mg        & 0.1460 & $-0.0008$ & 0.4782 \\
Ca        & $-0.1609$ & $-0.0041$ & 0.7836 \\
Sr        & $-0.1767$ & $-0.0201$ & 0.8312 \\
Li$^+$    & $\sim 0.0$ & $\sim 0.0$ & 0.0005 \\
Na$^+$    & 0.0003 & $\sim 0.0$ & 0.0004 \\
K$^+$     & 0.0004 & 0.0001 & 0.0070 \\
Rb$^+$    & 0.00960 & 0.0002 & 0.0115 \\
Sc$^{+}$  & $-0.0453$ & $-0.0533$ & 0.1874 \\
Y$^{+}$   & $-0.0631$ & $-0.0089$ & 0.4959 \\
Be$^{2+}$ & $\sim 0.0$ &  $\sim 0.0$ & 0.0002 \\
Mg$^{2+}$ & $0.0001$ &  $\sim 0.0$ & $0.0004$  \\
Ca$^{2+}$ & 0.0005 & 0.0001 & 0.0059 \\
Sr$^{2+}$ & 0.0061 & 0.0002 & 0.0052 \\
\end{tabular}
\end{ruledtabular}  
\end{table}  

With the knowledge of $T^{(0)}$ and $T^{(1)}$ amplitudes, we evaluate $\alpha$ using the relation
\begin{eqnarray}
\alpha &=&\langle\Phi_0 |e^{T^\dagger}D_Ne^{T}|\Phi_0 \rangle_{con} \nonumber \\
&=&\langle\Phi_0 |{T^{\dagger(1)}}\overbrace{D_N^{(0)}}+\overbrace{D_N^{(0)}}{T^{(1)}}|\Phi_0 \rangle_{con} \nonumber \\
&=&2 \langle\Phi_0 |\overbrace{D_N^{(0)}}(T_1^{(1)}+T_2^{(1)})|\Phi_0 \rangle_{con}
\end{eqnarray}
where $\overbrace{D_N^{(0)}} = e^{T^{\dagger{(0)}}}D_Ne^{T^{(0)}} $ is a non-truncating series. 
We compute this by dividing it into
connected effective one-body and two-body terms using the generalized Wick's theorem \cite{lindgren} before 
contracting them with $T^{(1)}$. In the CCSD approximation that we have considered here,
we need only the fully contracted terms from $\overbrace{D_N^{(0)}}(T_1^{(1)}+T_2^{(1)})$. Therefore only
the effective one-, two- and three-body terms will survive from $\overbrace{D_N^{(0)}}$.
Effective one-body diagrams arising from the non-truncating series $\overbrace{D_N^{(0)}}$ which contribute 
significantly are further binned into the hole-hole (H-H), particle-particle (P-P), hole-particle (H-P) and
particle-hole (P-H) type diagrams as shown in Fig. \ref{prpf0} (H-P diagrams are not shown as they are 
complex conjugate (cc) terms of P-H type of diagrams) considering up to minimum fifth order in the residual 
Coulomb interaction. It can be noticed from Fig. \ref{prpf0}(c) of the P-H and H-P type diagrams that they 
contain diagrams (e.g. $i, ii, iii$ etc.) resembling the random phase approximation (RPA) along with some of 
the non-RPA diagrams (e.g. $iv, v, vi$ etc.) which account for the core-polarization effects to all orders.
We found, as will be demonstrated in the next section, they are the leading contributors. Therefore, we 
replace the corresponding $D$ operator from
the P-H and H-P effective diagrams by the P-P and H-H diagrams as shown in Fig. \ref{prpf1} to dress-up the
effective H-P/P-H operators for evaluating these contributions more rigorously. These effective diagrams are 
then further combined with the $T^{(1)}$ and $T^{(0)}T^{(1)}$ operators to obtain the final contributions.
The important diagrams that make significant contributions from the effective two-body and three-body terms
of $\overbrace{D_N^{(0)}}$ are computed directly after contracting them with the $T^{(1)}$ operators.

We also store the effective H-P/P-H type one-body diagrams from $\overbrace{D_N^{(0)}}$ having more than two orders of residual Coulomb interaction, 
as shown in Fig. \ref{prpf2} separately and contract them with another set 
of $T^{(0)}T^{(1)}$ terms in the final calculations to include the contributions of core-polarization effects, higher in order
than those described above. We found that these contributions are crucial  for improving the final results  for the 
alkaline earth-metal atoms that were neglected in our previous calculations \cite{bijaya1,bijaya2,bijaya3}.
As they are just another set of effective one-body terms, they marginally affect the computational 
cost.
\begin{table}[t]
\caption{\label{tab0}
 Contributions to $\alpha$ at different level of approximations in
the many-body method.} 
\begin{ruledtabular}
\begin{tabular}{lcccc}
Atoms    & DF=MBPT(1)  & MBPT(2) & MBPT(3) & CCSD \\
\hline    \\
He  & 0.998   & 1.240  & 1.215  & 1.360 \\ 
Li$^+$  & 0.1579    & 0.1839  & 0.1851 & 0.1913 \\
Be$^{2+}$  & 0.0453  & 0.0510 & 0.0512 & 0.0521 \\

Ne  & 1.977   & 2.254  & 1.654  & 2.652 \\ 
Na$^+$  &  0.8337   & 0.9154  & 0.8504 & 0.9984 \\
Mg$^{2+}$  & 0.4277  & 0.4555  & 0.4371  & 0.4852 \\

Ar  & 10.152  & 9.964  & 8.005  & 11.089 \\
K$^+$   & 5.466      & 5.130  & 4.468  & 5.522 \\
Ca$^{2+}$  & 3.369   & 3.082  & 2.568  & 3.295 \\

Kr  & 15.82  & 15.00  & 10.70  & 16.93 \\
Rb$^+$  &  9.273    & 8.374  & 7.103  & 9.213 \\
Sr$^{2+}$   & 6.146  &  5.388  & 4.492 & 5.877 \\

Be  & 30.53   & 40.24  & 38.16 & 37.86 \\
Mg  & 54.70   & 70.72  & 65.64  & 72.54 \\
Ca  & 122.90  & 151.70  & 132.80 & 157.03 \\
Sc$^{+}$ & 50.10    & 57.17  & 47.02  & 53.24 \\
Sr  & 156.83  & 188.98  & 163.13 & 186.98 \\
Y$^{+}$  & 68.60    & 75.42  & 65.10  & 72.26 \\

\end{tabular}
\end{ruledtabular}
\end{table}        
\section{Results and Discussions}
We present our polarizability results for the ground states for several atomic systems
and compare them with other calculations 
and experimental results in Table \ref{tab1}. We have also estimated the errors in
our calculations arising from the numerical uncertainties due to the finite size of the basis,
neglected contributions from the Breit interaction and QED effects due to the lowest order
vacuum polarization and self-energy corrections. Moreover contributions from the 
Breit and QED effects are found to be small for the property under consideration here,
but the size of the basis is crucial for the numerical accuracy of the calculations. We estimated
contributions from the Breit and QED interactions using the MBPT(3) method by carrying out calculations
with these interactions separately along with the DC Hamiltonian. Inaccuracies from the choice of 
basis functions are estimated in two steps using the DF method: (i) results are obtained for different
set of optimized Gaussian parameters and (ii) estimating contributions from the inactive orbitals 
that are not considered in the RCC calculations from the DF method. We present these estimated
contributions from the individual source in Table \ref{tabx}. Experimental
results for light atomic systems are more accurate than our calculations. However,
for heavy systems, the accuracies of our results are better than those of experiments and 
many of the previous calculations.

\begin{figure}[t]
\includegraphics[width=7cm, height=4cm]{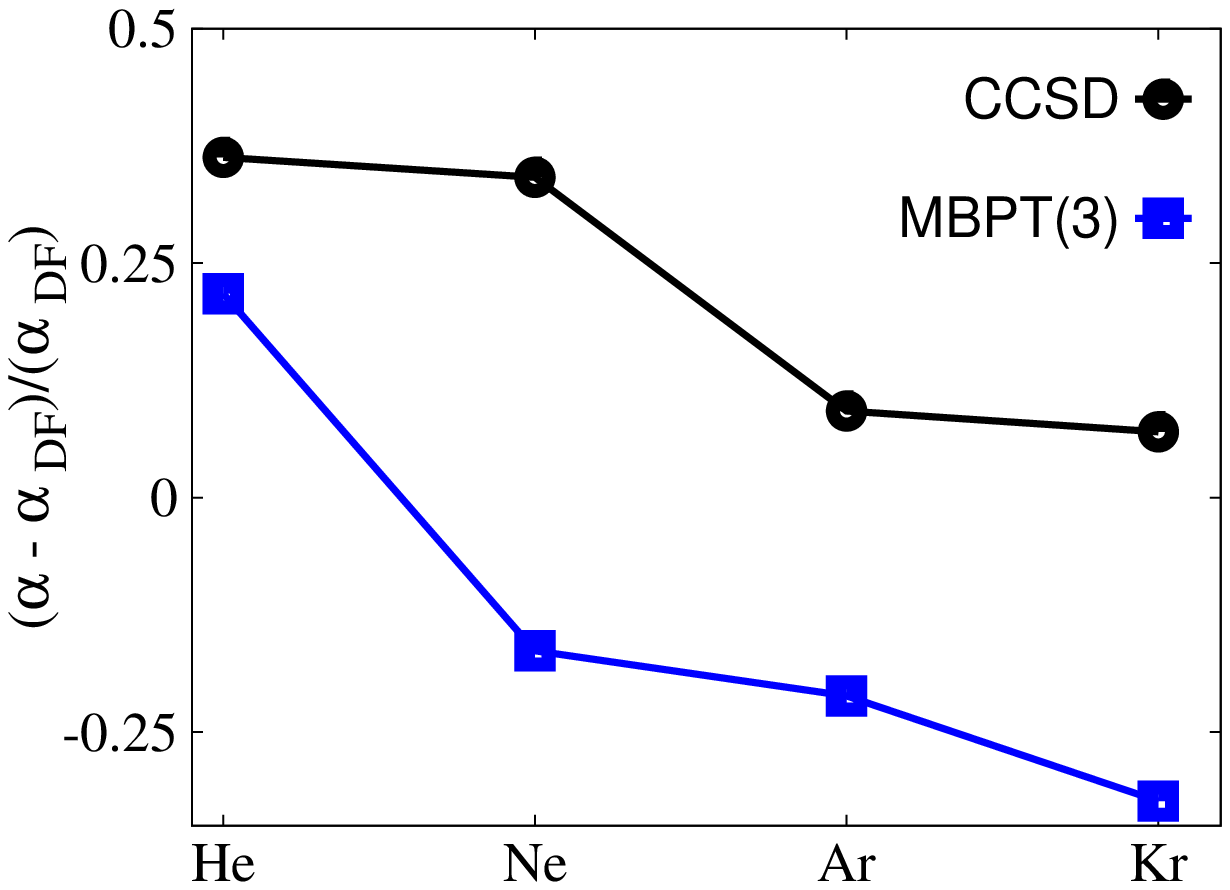}
 \includegraphics[width=7cm, height=4cm]{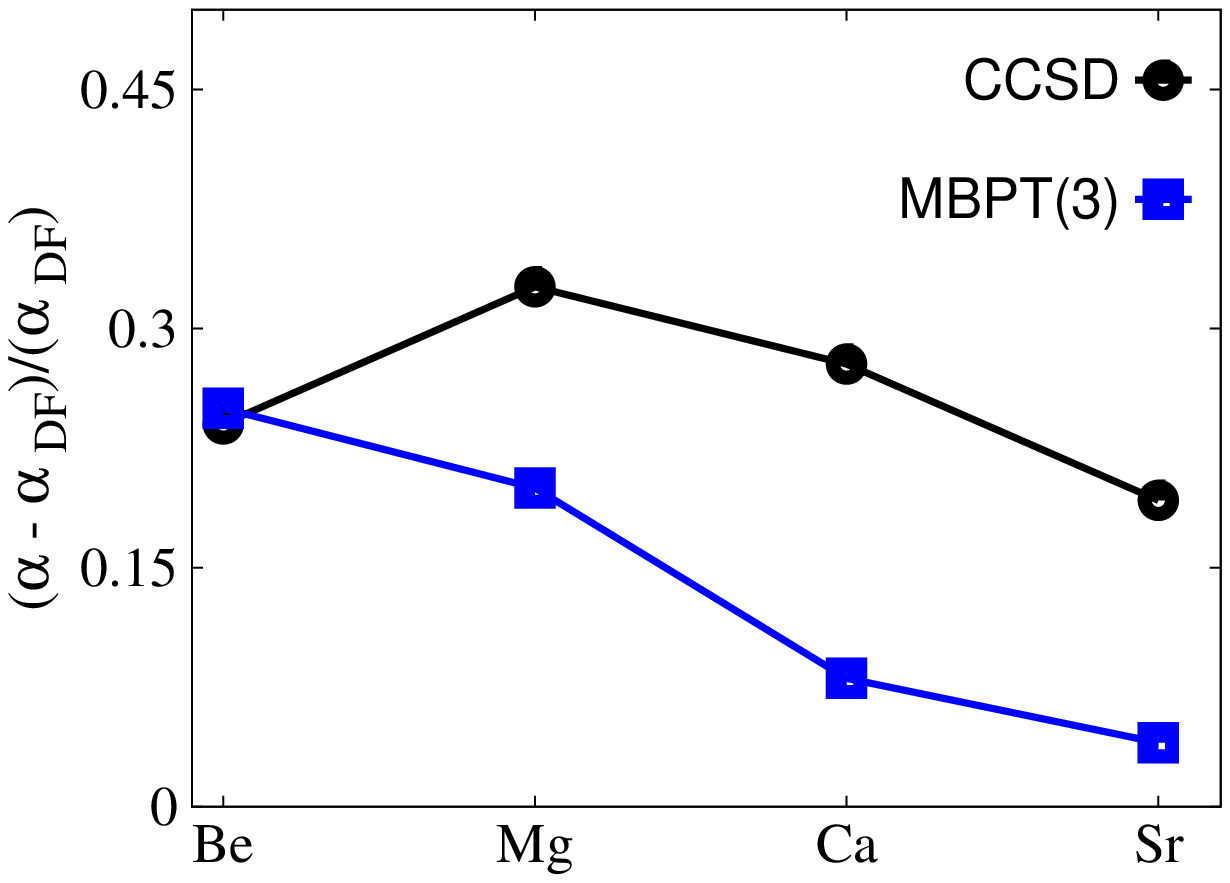}
\includegraphics[width=7cm, height=4cm]{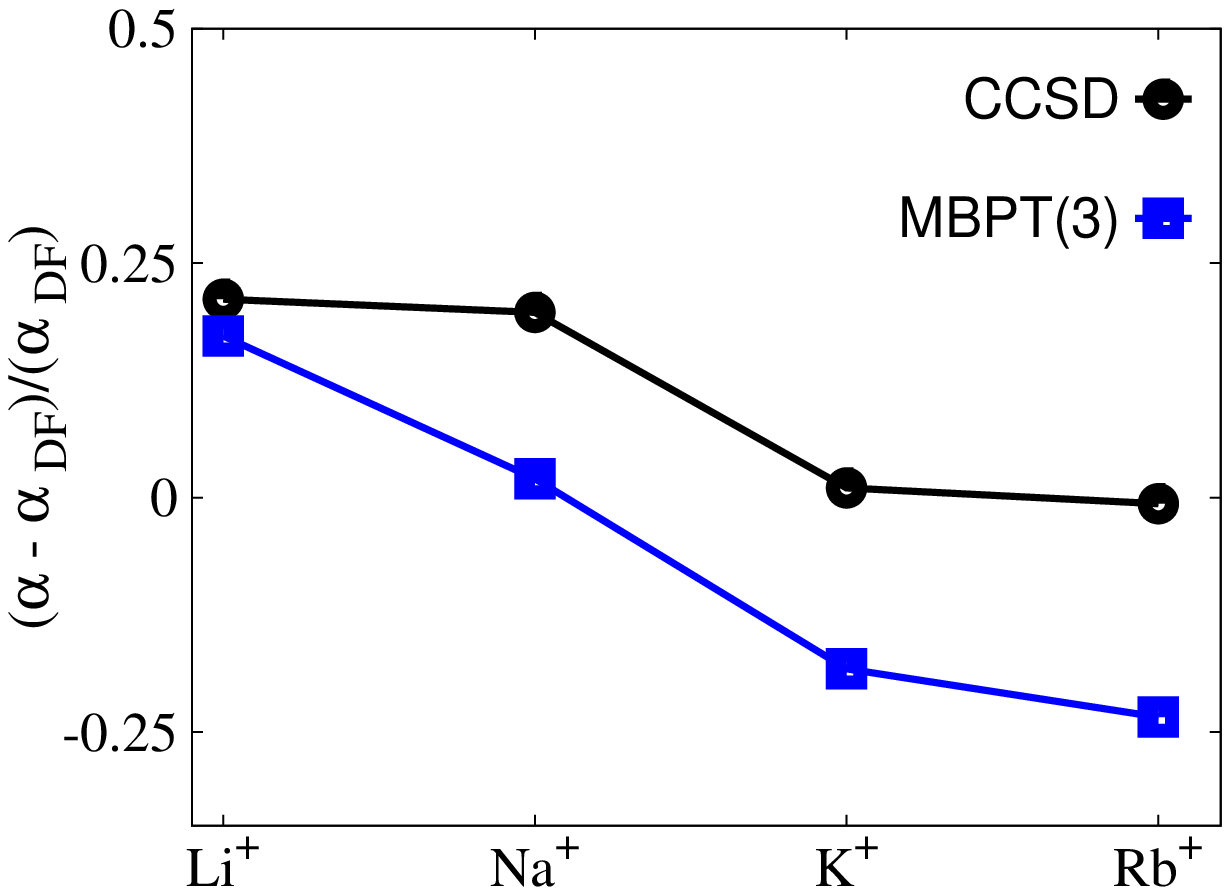}
\includegraphics[width=7cm, height=4cm]{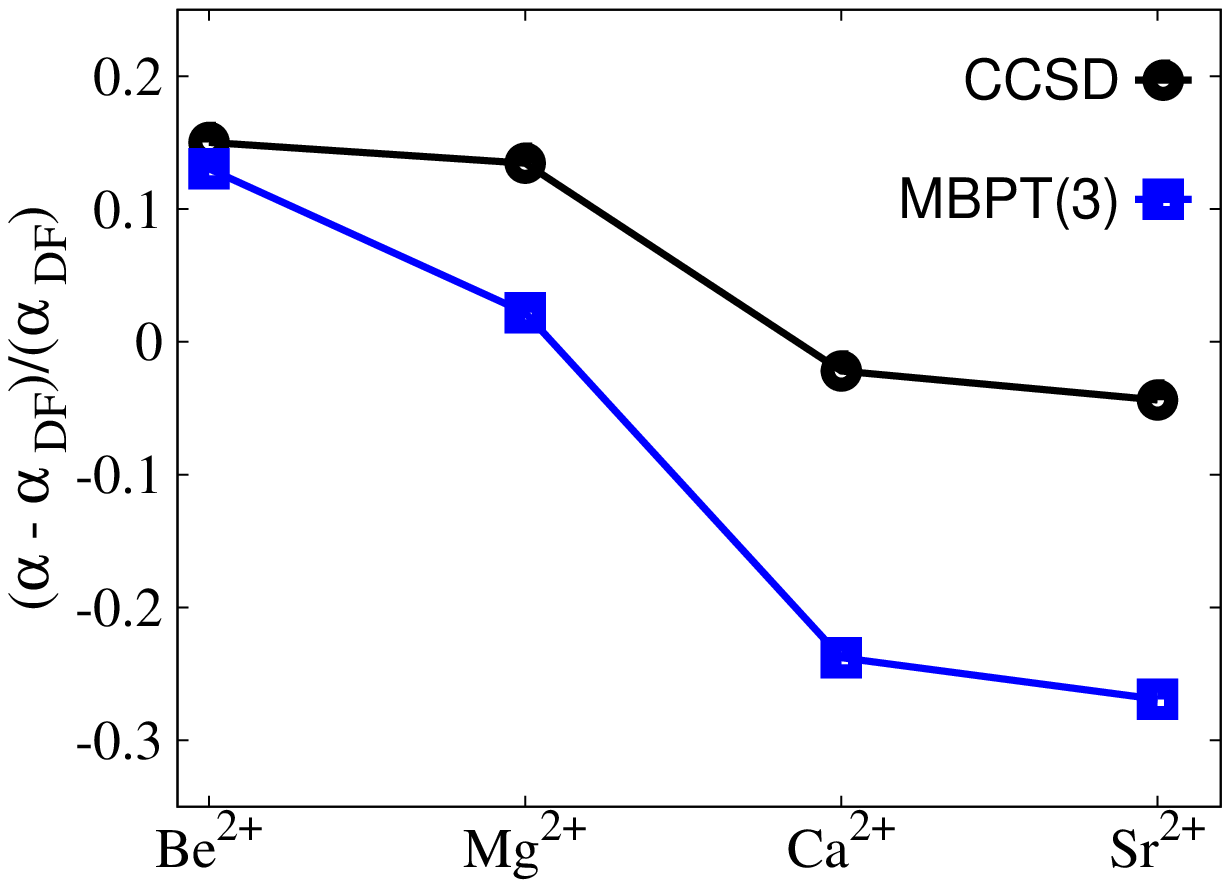}
\caption{(color online) Plots of $(\alpha - \alpha_{DF})/\alpha_{DF}$ results versus atomic numbers from
different groups of atomic systems. $\alpha$ results are obtained using the MBPT(3)
and CCSD methods in order to make a comparative study between these two approaches.}
\label{cocf}
\end{figure}
\begin{table}[t]
\caption{\label{tabm}
Contributions from the RPA and non-RPA type of diagrams from the MBPT(3) method.}
\begin{ruledtabular}
\begin{tabular}{lccc}
\textrm{System}& RPA & non-RPA  \\
\hline                                                                                      \\
He        & 1.274  & $-0.059$     \\
Ne        & 2.303  & $-0.649$     \\
Ar        & 9.878  & $-1.873$     \\
Kr        & 14.980 &  $-4.280$    \\
Be        & 36.788 &  1.372   \\
Mg        & 65.074 &  0.566    \\
Ca        & 135.459&  $-2.659$    \\
Sr        & 170.340&  $-7.210$    \\
Li$^+$    & 0.1862 &   $-0.0011$   \\
Na$^+$    & 0.9261 &   $-0.0757$   \\
K$^+$     & 5.035  &    $-0.567$  \\
Rb$^+$    & 8.326        &    $-1.223$  \\
Sc$^{+}$  & 50.115        &   $-3.095$   \\
Y$^{+}$   & 67.181        &   $-2.081$   \\
Be$^{2+}$ & 0.0513        &   $-0.0001$   \\
Mg$^{2+}$ & 0.4627        &   $-0.0256$   \\
Ca$^{2+}$ & 3.009        &    $-0.441$  \\
Sr$^{2+}$ & 5.352        &    $-0.860$  \\
\end{tabular}
\end{ruledtabular}  
\end{table}  

A variety of many-body methods have been used to determine $\alpha$ for the systems that we have considered
except for Sc$^{+}$ and Y$^{+}$. The method we have employed in the present work 
had been used previously to calculate these quantities \cite{bijaya1,bijaya2,bijaya3}. In those calculations,
we had truncated $\overbrace{D_N^{(0)}}$ at $\overbrace{D_N^{(0)}}=T^{\dagger(0)}D_NT^{(0)}$ neglecting
higher order RPA contributions coming through the $T^{\dagger(0)}D_N(T^{(0)})^2$ and 
$(T^{\dagger(0)})^2D_NT^{(0)}$ whose contributions are found to be significant in the neutral alkaline
earth atoms, Sc$^+$ and Y$^+$. Also the results have improved, particularly in these systems, since 
the dipole operators in the P-H and H-P effective diagrams described by Fig. \ref{prpf2} and in the
construction of the effective two-body operators have been used. Recently a similar approach,
which had included the normalization of the wave function, had been used for evaluating the $\alpha$'s of
some of the inert gas atoms \cite{sidhu-ne, sidhu-nobel}. In fact, both these works account for non-linear
terms at different level of approximations resulting in some differences in the results. Another calculation
for the inert gas atoms was carried out by Nakajima and Hirao \cite{nakajima}, where they have
investigated the relativistic effects in $\alpha$ using a scalar relativistic Douglas-Kroll (DK) Hamiltonian.
The other difference between this work and ours is that Nakajima and Hirao had estimated polarizability
from the second order energy shift due to an arbitrary external electric field by a numerical finite field
approach whereas we have 
evaluated this quantity by calculating the expectation value of the dipole operator using the first order dipole perturbed wave function. It
is interesting that both the results agree fairly well with each other within the quoted
uncertainties.
\begin{figure*}[t]
\includegraphics[width=14cm, height=4cm]{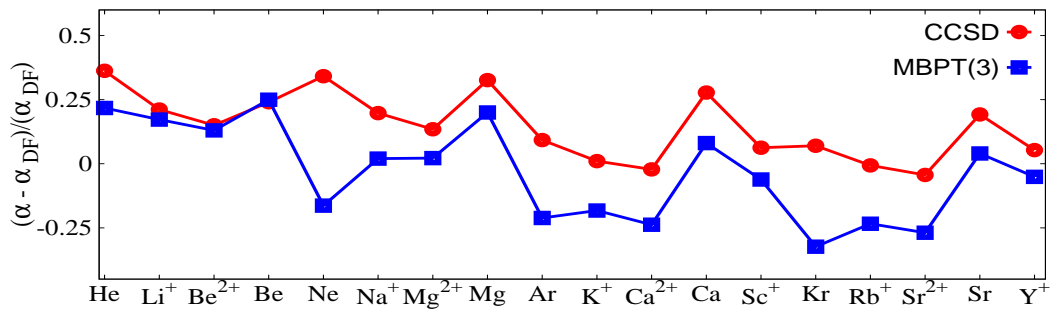}
\caption{(color online) Plots of MBPT(3) and CCSD $(\alpha \textendash \alpha_{DF} )/\alpha_{DF}$ results for all the considered atoms and ions versus atomic numbers.}
\label{cocfa}
\end{figure*} 
For the alkaline earth elements we compare our results of $\alpha$ with those of Porsev {\it et al.} \cite{porsev1,porsev2},
who had used a hybrid approach combining the configuration interaction (CI) method in the valence 
space and the MBPT method by scaling the energies and dressing the external electromagnetic 
field in the RPA framework to evaluate the core-polarization effects. Lim {\it et. al} \cite{lim} had performed the
RCC calculations in the finite field method using the DK Hamiltonian. Our results for most of the atoms are in good agreement 
with them except for Sr which differs significantly. As mentioned in the previous section, we have found that higher
order non-linear terms, especially those corresponding to RPA, are essential for obtaining accurate results. One probable reason 
for the discrepancies in 
the results between our RCC results and those of Porsev {\it et al.} is the treatment of core-core correlation
effects in the two cases. We have computed these effects by the all order CCSD method, while they have used a finite
order MBPT approach. An important difference between our approach and that of Lim {\it et. al} \cite{lim} is that we have used
the proper DC Hamiltonian unlike its scalar components in \cite{lim} and the polarizabilities are estimated from the second  
order corrections to their calculated energies. 

We now compare our results for the singly
charged alkali-metal and doubly charged alkaline-earth-metal ions, which have electronic configurations
similar to that of the inert gas atoms with the results obtained using the RPA method by Johnson {\it et. al.} \cite{johnson1}
and another RCC calculation \cite{sidhu-alkali,sidhu-alkaline}. The method employed in the latter calculations 
have already been referred to in the previous paragraph. The close agreement between the RPA and our RCC results 
is due to the fact that the dominant correlation effects in the evaluation of the polarizabilities for the 
closed-shell systems come from the core-polarization effects which are taken to all orders in both the calculations.
We find from the MBPT(3) calculations that the non-RPA diagrams also contribute significantly in the 
closed-shell atoms. However, they cancel out to a large extent in these ions and their net contributions are consequently
not significant.
In another work, Lim {\it et. al} \cite{lim2} have reported results for the alkali 
ions considering the scalar relativistic DK Hamiltonian and accounting for the spin-orbit coupling corrections through
the MBPT(2) method using a fully relativistic four-component DF wave functions.
In addition to the above mentioned systems we have also calculated $\alpha$ for Sc$^+$ and Y$^+$ ions, 
but there are no data available for comparison with our results. 

\begin{table}[t]
\caption{\label{tab2}
Contributions from various CCSD terms for the evaluation of $\alpha$ in the ground states
of the considered atomic systems.}
\begin{ruledtabular}
\begin{tabular}{llllll}
\textrm{System}& \textrm{$DT_1^{(1)}$}& \textrm{$T_1^{(0)\dagger}DT_1^{(1)}$}& \textrm{$T_2^{(0)\dagger}DT_1^{(1)}$}& \textrm{$T_2^{(0)\dagger}DT_2^{(1)}$}& \textrm{Others} \\
\hline                                                                                      \\
He             & 1.362  &  0.005  & $-$0.035  & 0.035  & $-$0.007  \\

Ne             & 2.613  &  0.073  & $-$0.099  & 0.089  & $-$0.024  \\

Ar             & 11.806 & $-$0.068  & $-$1.143   & 0.511  & $-$0.017  \\

Kr             & 18.11  & $-$1.12   & $-$1.82    & 0.74  &  1.02  \\

Be             & 39.45 & $-$1.53  & $-$7.21    & 3.84  & 3.31   \\

Mg             & 75.66 & $-$2.96  & $-$10.16    & 5.54  & 4.46   \\

Ca             & 163.87& $-$9.24  & $-$24.89   & 16.05 & 11.24  \\

Sr             & 201.90  & $-$12.77 & $-$28.77  & 15.57  &  11.05 \\

Li$^+$         & 0.1894 & $\sim$0 & 0.0019   & 0.0019 & $-$0.0019 \\

Na$^+$         & 0.9756 & $\sim$0 & $-$0.0005   & $\sim$0 & 0.0233 \\

K$^+$          & 5.972  & $-$0.038  & $-$0.620   & 0.211   & $-$0.003  \\

Rb$^+$         & 9.971  & $-$0.067  & $-$1.049   & 0.333   & 0.025  \\

Sc$^{+}$       & 61.71 & $-$2.16  & $-$8.24   & 3.84   & $-$1.91  \\

Y$^{+}$        & 83.19 & $-$3.18  & $-$10.68  & 5.05   & $-$2.12   \\

Be$^{2+}$      & 0.0526 & $\sim$0 & $-$0.0007  & 0.0003  & $-$0.0001 \\

Mg$^{2+}$      & 0.4774 & $\sim$0 & $\sim$0 & $\sim$0 & 0.0078 \\

Ca$^{2+}$      & 3.578  & $-$0.019  & $-$0.585   & 0.117   & 0.204  \\

Sr$^{2+}$      & 6.396  & $-$0.037  & $-$0.689   & 0.191   & 0.016  \\
               
\end{tabular}
\end{ruledtabular}  
\end{table}  
The main aim of the present work is to analyze the trends in the correlation effects in the
static electric dipole polarizabilities of the ground states in a variety of closed shell atomic systems 
evaluated by different many-body methods in order to assess their potential for yielding accurate results
for the coupling constants associated with the permanent electric dipole 
moments due to parity and time-reversal violations in atoms of experimental interest \cite{rosenberry-xe,guest-ra,romalis-hg,griffith-hg,furukawa-xe,rand-rn,inoue-xe}.
To fulfill our objective, we have carried out a range of calculations using lower order MBPT to MBPT(3) methods and have presented
the results at each stage in Table \ref{tab0}. These results are further compared with our final
all order calculation using the relativistic CCSD method in the same table. This clearly demonstrates the importance of the correlation effects
starting from lower to higher order perturbation theory systematically and provides a good understanding of their roles in obtaining accurate results.
We have given results by grouping the iso-electronic systems together in this table in order to make a comparative analysis of the correlation 
effects as the atomic number varies. As can be seen in the table, the DF results are smaller than the MBPT results
for the light inert gas atoms while this trend changes for the heavier ones. Finally,
the CCSD results are larger than the DF results for all these atoms. Therefore this implies that there are 
strong cancellations between the correlation effects in these atoms and the higher order correlation effects
play a pivotal role in determining the final results. A similar trend is also followed by other inert gas like
singly charged alkali and doubly charged alkaline earth-metal ions. However, the trend for the 
correlation effects in the neutral alkaline earth-metal atoms is rather different. In this case the DF results are always smaller than those of  
MBPT and CCSD. In fact, it is also quite interesting to note that the correlation trends for Sc$^+$ and Y$^+$ do not follow 
those of other iso-electronic alkaline earth-metal atoms, but rather of the inert gas atoms. 
For a quantitative description, we plot $(\alpha - \alpha_{DF})/\alpha_{DF}$ obtained using the MBPT(3) 
and CCSD methods versus atomic numbers in in Fig. \ref{cocf} for the different categories of systems that we have considered. We
also plot the same for all the systems together including Sc$^+$ and Y$^+$ ions in Fig. \ref{cocfa}
to make a comparative analysis of the correlation trends among different iso-electronic sequences.
To shed light on the role of different types of correlation effects that are crucial in the determination of
polarizabilities and to explain the reasons for their trends in different iso-electronic sequences,
we identify diagrams from the MBPT(3) approximation that belong to lower order RPA. We, then, present MBPT(3) results
in Table \ref{tabm} classifying its diagrams as RPA and non-RPA types. As can be seen from Fig. \ref{fig1} that 
all the diagrams up to MBPT(2) belong to RPA and hence, they are the dominant contributors.
However, diagrams shown in Fig. \ref{fig1}(vi-viii) are few examples of non-RPA type diagrams that also 
contribute significantly at the third order level, but they largely cancel out each other in the heavy atomic systems. The final
results are the outcome of the interplay between these cancellations which can only be accounted
correctly using an all order method like our CCSD method. This is evident from the contributions
of the separate RCC terms presented below.
 
We now present the contributions from different correlation effects represented by the RCC terms in the
evaluation of $\alpha$ for different atomic systems. In Table \ref{tab2}, we give the individual contribution
from the important CCSD terms to $\alpha$, where the leading term $DT_1^{(1)}$ contains the lowest order DF result. 
The next important term is $T_2^{(0)\dagger}DT_1^{(1)}$ and the sign of its contribution is opposite to that of the former resulting in a substantial 
cancellation between these two largest contributors. In addition to the above two terms, contributions from $T_1^{(0)\dagger}DT_1^{(1)}$ further
reduce the final results. Pair excitations contributing through $T_2^{(0)\dagger}DT_1^{(1)}$ and other
higher order non-linear terms together take our final results towards the experimental values.

\section{Conclusion}
We have employed the relativistic coupled-cluster method to calculate the static electric dipole polarizabilities
of the ground states of many closed shell atomic systems. We have improved the results of our previous calculations
for some of them by adding important non-linear RCC terms which correspond to higher order correlation effects in the 
present calculations. The crucial role of correlation effects is highlighted by presenting and comparing the results at different levels
of approximations from lower order many-body perturbation theory to the relativistic CCSD method. 
Correlation trends among the neutral atoms, singly charged ions and doubly charged ions are presented.
Investigation of various correlation effects in evaluating polarizabilities will provide valuable insights into the ongoing theoretical 
work on atomic electric dipole moments which arises due to parity and time-reversal symmetry violation. Our results will also serve as a guide to the
future measurements of systems where the experimental values of polarizabilities are not precisely known.
\section{Acknowledgment}
We thank D. K. Nandy for his partial contribution in few parts of our programming. 
BKS and BPD were supported partly by INSA-JSPS under project no. 
IA/INSA-JSPS Project/2013-2016/February 28, 2013/4098.
The computations reported in the present work were carried out using the 3TFLOP HPC cluster at
Physical Research Laboratory, Ahmedabad.

\end{document}